\documentclass[twocolumn,english,american]{revtex4-2}
\usepackage[T1]{fontenc}
\usepackage[utf8]{luainputenc}
\usepackage{amsmath}
\usepackage{amssymb}
\usepackage{babel}
\begin{document}
\title{One Axis Twisting (OAT) spin squeezing for metrology}
\author{Garry Goldstein$^{1}$}
\address{$^{1}$garrygoldsteinwinnipeg@gmail.com}
\begin{abstract}
In this work we study One Axis Twisting (OAT) spin squeezing for metrology
in the presence of decoherence. We study Linbladian evolution in the
presence of both $T_{1}$ and $T_{2}$ (longitudinal and transverse
relaxation processes). We show that spin squeezing can be an effective
way to improve metrological accuracy even in the presence of decoherence
for OAT squeezing. We show our results are not sensitive to inhomogeneity
of the squeezing strength of the many spin OAT Hamiltonian and that
very general squeezed states do not have entanglement enhanced decoherence.
We also extend the Kitagawa-Ueda OAT squeezing formula to finite polarization
$P$.
\end{abstract}
\maketitle

\section{Introduction}\label{sec:Introduction}

In the past thirty years spin squeezing has been one of the focal
points of study of both experimental and theoretical physics \citep{Kitagawa_1993,Wineland_1992,Wineland_1994}.
Spin squeezing can be used to study correlations and entanglement
in many particle systems as well as to improve the precision of measurements
of external fields say due to Ramsey interferometry. Indeed spin squeezing
can be shown to be an entanglement witness where if the squeezing
parameter $\xi^{2}<1$ along some axis the state can be shown to be
an entangled state of spins \citep{Sorensen_2001,Bigelow_2001,Guehne_2009,Polzik_2008,Cronin_2009}.
Furthermore the precision of Ramsey magnetometry measurements can
be improved for squeezed states. Indeed the squeezed direction for
the state has lesser variance then that of coherent states and can
be used to improve both the speed and precision of measurement by
decreasing the quantum noise of the measurement. Spin squeezing can
be generated in a variety of ways including atom atom interactions
in two component BEC's and induced spin spin interactions in a far
detuned Dicke model \citep{Ma_2011}. The notion that OAT spin squeezing
can be used to enhance metrological precision in the absence of decoherence
is well established however the situation is more complex \citep{Kitagawa_1993,Wineland_1992,Wineland_1994}.
Some studies of this problem focusing on squeezing in the presence
of atom loss for two component BECs \citep{Li_2008,Li_2009} and studying
the effects of decoherence after OAT squeezing \citep{Ma_2011} showing
this to be a promising direction. In this work we will study spin
interacting with an OAT Hamiltonian in the presence of generic decoherence
(see Eq. (\ref{eq:Linbladian})). We will show analytically that squeezing
under decoherence is equivalent to squeezing with reduced initial
polarization. We study squeezing for reduced initial polarization
(see Appendix \ref{sec:Finite-polarization-effects}) and find its
effects on maximal attainable spin squeezing (see Eq. (\ref{eq:Squeezing_final})).
We also study metrology for squeezed states and find that metrology
with squeezed states under decoherence is equivalent to metrology
with a reduced external field and reduced initial polarization (see
Eq. (\ref{eq:Fields})). We find the maximal sensitivity achievable
for this setup is given by Eq. (\ref{eq:Signal_noise}). In Appendix
\ref{sec:Varaible-couplign-squeezing} we show that squeezing is virtually
insensitive to reasonable inhomogeneities in the squeezing strength
Hamiltonian and in Appendix \ref{sec:Decay-of-squeezed} we show that
very general squeezed states are not strongly affected by decoherence,
that is there is no entanglement enhanced decoherence rate for them. 

\section{Squeezing under decoherence}\label{sec:Squeezing-under-decoherence}

\subsection{General Setup}\label{subsec:General-Setup}

We consider a spin ensemble with $N$ spins with the initial density
matrix given by 
\begin{equation}
\rho=\otimes_{i=1}^{N}\left(\frac{\mathbb{I}}{2}+P\sigma_{z}^{i}\right)\label{eq:Density}
\end{equation}
Here $\sigma^{i}$ are the Pauli matrices. We consider the following
Linblad evolution for the system
\begin{align}
\frac{\partial\rho}{\partial t} & =\mathcal{L}_{1}\left(\rho\right)+\mathcal{L}_{2}\left(\rho\right)\nonumber \\
\mathcal{L}_{1}\left(\rho\right) & =-i\left[H_{Squ},\rho\right]\nonumber \\
\mathcal{L}_{2}\left(\rho\right) & =\Gamma_{\parallel}\sum_{i}\sigma_{x}^{i}\rho\sigma_{x}^{i}+\Gamma_{\perp}\sum_{i}\left[\sigma_{z}^{i}\rho\sigma_{z}^{i}+\sigma_{y}^{i}\rho\sigma_{y}^{i}\right]\nonumber \\
 & \qquad-\left(\Gamma_{\parallel}+2\Gamma_{\perp}\right)\rho\nonumber \\
H_{Squ} & =J\sum_{i,j}\sigma_{x}^{i}\sigma_{x}^{j}\label{eq:Linbladian}
\end{align}
We now see that after time $T$:
\begin{align}
\rho\left(T\right) & =\exp\left[T\left(\mathcal{L}_{1}+\mathcal{L}_{2}\right)\right]\rho\nonumber \\
 & =\exp\left[T\mathcal{L}_{1}\right]\exp\left[T\mathcal{L}_{2}\right]\rho\label{eq:Factorization}
\end{align}
 Where we have used that:
\begin{equation}
\left[\mathcal{L}_{1},\mathcal{L}_{2}\right]=0+O\left(\frac{1}{N}\right)\label{eq:Commutator_zero}
\end{equation}
Which can be checked directly or qualitatively as we note that for
both Linbladians $\mathcal{L}_{1}$ and $\mathcal{L}_{2}$ only the
x-axis is singled out and treat the rest of the spin density matrix
symmetrically. Indeed one represents rotations about the x-axis and
the other decrease of the polarization with the x-axis singled out.
As such rhe two evolutions almost commute (see Appendix \ref{sec:Argument-why-Eq.}).
Now we see that 
\begin{equation}
\exp\left[T\mathcal{L}_{2}\right]\rho=\otimes_{i=1}^{N}\left(\frac{\mathbb{I}}{2}+P\exp\left[-2\left[\Gamma_{\perp}+\Gamma_{\parallel}\right]T\right]\sigma_{z}^{i}\right)\label{eq:Loss_phase}
\end{equation}
As such we have reduced the squeezing under decoherence problem to
a pure squeezing problem with finite initial polarization.

\subsection{Calculation for squeezing}\label{subsec:Calculation-for-squeezing}

In Appendix \ref{sec:Finite-polarization-effects} we derive that
for the initial state with polarization $\mathcal{P}=P\exp\left[-2\left[\Gamma_{\perp}+\Gamma_{\parallel}\right]T\right]$
the optimum squeezing is given by:
\begin{align}
 & \xi_{min}^{2}\left(T\right)=P^{-1}\exp\left[2\left[\Gamma_{\perp}+\Gamma_{\parallel}\right]T\right]\times\nonumber \\
 & \times\left[\frac{P^{-2}\exp\left[4\left[\Gamma_{\perp}+\Gamma_{\parallel}\right]T\right]}{16N^{2}J^{2}T^{2}}+\frac{32}{3}N^{2}J^{4}T^{4}\right]\label{eq:Minimal_squeezing}
\end{align}
Where $\xi_{min}^{2}\left(T\right)$ is the squeezed quadrature (minimal
variance quadrature). Now let us introduce 
\begin{equation}
\Theta=2\left[\Gamma_{\perp}+\Gamma_{\parallel}\right]T\label{eq:variable}
\end{equation}
We find that: 
\begin{align}
 & \xi_{min}^{2}\left(T\right)=P^{-1}\exp\left[\Theta\right]\times\nonumber \\
 & \times\left[\frac{P^{-2}\left[\Gamma_{\perp}+\Gamma_{\parallel}\right]^{2}\exp\left[2\Theta\right]}{4N^{2}J^{2}\Theta^{2}}+\frac{2N^{2}J^{4}\Theta^{4}}{3\left[\Gamma_{\perp}+\Gamma_{\parallel}\right]^{4}}\right]\label{eq:Intermediate_answer}
\end{align}
While it is possible to optimize this expression numerically as a
function of $\Theta$ in many cases decoherence is more important
then over squeezing in Eq. (\ref{eq:Intermediate_answer}) in which
case its sufficient to optimize: 
\begin{equation}
\frac{\partial}{\partial\Theta}\frac{\exp\left(3\Theta\right)}{\Theta^{2}}=0\Rightarrow\Theta_{min}=\frac{2}{3}\label{eq:Minimum}
\end{equation}
In which case 
\begin{align}
 & \xi_{min}^{2}\cong1.948P^{-1}\times\nonumber \\
 & \times\left[2.134\frac{P^{-2}\left[\Gamma_{\perp}+\Gamma_{\parallel}\right]^{2}}{4N^{2}J^{2}}+0.132\frac{N^{2}J^{4}}{\left[\Gamma_{\perp}+\Gamma_{\parallel}\right]^{4}}\right]\label{eq:Squeezing_final}
\end{align}

\section{Metrology for squeezing under decoherence}\label{sec:Metrology-for-squeezing}

We again consider a spin ensemble with $N$ spins with the initial
density matrix given by Eq. (\ref{eq:Density}). We consider the following
Linblad evolution:
\begin{align}
\frac{\partial\rho}{\partial t} & =\mathcal{L}_{1}\left(\rho\right)+\mathcal{L}_{2}\left(\rho\right)+\mathcal{L}_{3}\left(\rho\right)\nonumber \\
\mathcal{L}_{1}\left(\rho\right) & =-i\left[H_{Squ},\rho\right]\nonumber \\
\mathcal{L}_{2}\left(\rho\right) & =\Gamma_{\parallel}\sum_{i}\sigma_{x}^{i}\rho\sigma_{x}^{i}+\Gamma_{\perp}\sum_{i}\left[\sigma_{z}^{i}\rho\sigma_{z}^{i}+\sigma_{y}^{i}\rho\sigma_{y}^{i}\right]\nonumber \\
 & \qquad-\left(\Gamma_{\parallel}+2\Gamma_{\perp}\right)\rho\nonumber \\
\mathcal{L}_{3}\left(\rho\right) & =-i\left[H_{Sig},\rho\right]\nonumber \\
H_{Squ} & =J\left(t\right)\sum_{i,j}\sigma_{x}^{i}\sigma_{x}^{j}\nonumber \\
H_{Sig} & =\sum_{i}B_{y}\sigma_{y}^{i}\label{eq:Linblad_evolution}
\end{align}
Here for simplicity we will assume that $J\left(t\right)$ is turned
on and off rapidly on the scale of the experiments and treat $B_{y}$
as small working only to linear oder in $B_{y}$. We now see that:
\begin{align}
\rho\left(T\right) & =\exp\left[T.O.\int_{0}^{T}dt\left(\mathcal{L}_{1}\left(t\right)+\mathcal{L}_{2}\left(t\right)+\mathcal{L}_{3}\left(t\right)\right)\right]\rho\nonumber \\
 & \cong\left[\mathbb{I}+\int_{0}^{T}dt\exp\left(t\mathcal{L}_{2}\right)\mathcal{L}_{3}\exp\left(-t\mathcal{L}_{2}\right)\right]\times\nonumber \\
 & \times\exp\left[\int_{0}^{T}dt\mathcal{L}_{1}\left(t\right)\right]\exp\left[T\mathcal{L}_{2}\right]\rho\label{eq:Linear_response-1}
\end{align}
Where $T.O.$ stands for time ordering and we went to the master equation
interaction picture and kept only linear terms in $B_{y}$. Now we
have that:
\begin{align}
 & \exp\left(t\mathcal{L}_{2}\right)\mathcal{L}_{3}\exp\left(-t\mathcal{L}_{2}\right)\rho=\nonumber \\
 & -i\left[\sum_{i}B_{y}\sigma_{y}^{i}\exp\left[-2\left[\Gamma_{\perp}+\Gamma_{\parallel}\right]T\right],\rho\right]\label{eq:Decay_field}
\end{align}
 This means that 

\begin{align}
 & \left[\mathbb{I}+\int_{0}^{T}dt\exp\left(t\mathcal{L}_{2}\right)\mathcal{L}_{3}\exp\left(-t\mathcal{L}_{2}\right)\right]\times\nonumber \\
 & \times\exp\left[\int_{0}^{T}dt\mathcal{L}_{1}\left(t\right)\right]\exp\left[T\mathcal{L}_{2}\right]\rho\nonumber \\
 & \cong\exp\left(-i\sum_{i}\mathfrak{B}\sigma_{y}^{i}\right)\left[\exp\left[\int_{0}^{T}dt\mathcal{L}_{1}\left(t\right)\right]\exp\left[T\mathcal{L}_{2}\right]\rho\right]\times\nonumber \\
 & \times\exp\left(i\sum_{i}\mathfrak{B}\sigma_{y}^{i}\right)\label{eq:Linear_response}
\end{align}
Where 
\begin{equation}
\mathfrak{B}=\frac{B_{y}}{2\left[\Gamma_{\perp}+\Gamma_{\parallel}\right]}\left(1-\exp\left(-2\left[\Gamma_{\perp}+\Gamma_{\parallel}\right]T\right)\right)\label{eq:Fields}
\end{equation}
From this we see that when we measure along the axis of minimal quadrature
(which is close to the x-axis so we may ignore some geometric angle
effects (see Appendix \ref{sec:Finite-polarization-effects}) the
signal to noise is given by: 
\begin{align}
 & \frac{Signal}{Noise}=\sqrt{\frac{\tau}{T}}\frac{B_{y}}{2\left[\Gamma_{\perp}+\Gamma_{\parallel}\right]}\times\nonumber \\
 & \times\frac{\left(1-\exp\left[-2\left[\Gamma_{\perp}+\Gamma_{\parallel}\right]T\right]\right)P\exp\left[-2\left[\Gamma_{\perp}+\Gamma_{\parallel}\right]T\right]}{\left[\frac{P^{-2}\exp\left[4\left[\Gamma_{\perp}+\Gamma_{\parallel}\right]T\right]}{16N^{2}J^{2}T^{2}}+\frac{32}{3}N^{2}J^{4}T^{4}\right]}\label{eq:Signal}
\end{align}
Here $\tau$ is the total measurement time and we have defined $J=\frac{1}{T}\int J\left(t\right)dt$.
Then we obtain that: 
\begin{align}
 & \frac{\partial\mathcal{S}\left(\Theta\right)}{\sqrt{\tau}\partial B_{y}}=\frac{2^{3/2}N^{2}J^{2}P^{3}}{\left[\Gamma_{\parallel}+\Gamma_{\perp}\right]^{5/2}}\times\nonumber \\
 & \times\frac{\Theta^{3/2}\exp\left[-3\Theta\right]\left[1-\exp\left[-\Theta\right]\right]}{\left[1+\frac{2}{3\left[\Gamma_{\parallel}+\Gamma_{\perp}\right]^{6}}P^{2}N^{4}J^{6}\Theta^{6}\exp\left(-2\Theta\right)\right]}\label{eq:Dimensionless}
\end{align}
While it is possible to to optimize this expression numerically decoherence
often dominates over over-squeezing for OAT and its sufficient to
optimize: 
\begin{equation}
\frac{\partial}{\partial\Theta}\Theta^{3/2}\exp\left[-3\Theta\right]\left[1-\exp\left[-\Theta\right]\right]=0\Rightarrow\Theta_{max}\cong0.727\label{eq:Optimize_numeric}
\end{equation}
This means that: 
\begin{equation}
\frac{\partial\mathcal{S}_{max}}{\sqrt{\tau}\partial B_{y}}\cong\frac{N^{2}J^{2}P^{3}}{\left[\Gamma_{\parallel}+\Gamma_{\perp}\right]^{5/2}}\frac{0.205}{\left[1+\frac{0.092}{\left[\Gamma_{\parallel}+\Gamma_{\perp}\right]^{6}}P^{2}N^{4}J^{6}\right]}\label{eq:Signal_noise}
\end{equation}

\section{Conclusions}\label{sec:Conclusions}

In this work we have studied squeezing under decoherence. We have
shown metrological gain for squeezing in the presence of decoherence.
In the Appendix we show this result to be robust to inhomogeneities
in the the squeezing parameter and to partial initial polarizations.
In the appendix we have also shown that very general squeezed states
are not strongly affected by dephasing in that there is no entanglement
enhanced decoherence. 
\selectlanguage{english}%

\appendix

\section{Decay of squeezed states due to dephasing}\label{sec:Decay-of-squeezed}

In this appendix we show that generic squeezed states are not sensitive
to decoherence. We will consider a generic squeezed state, not just
the one given by Eq. (\ref{eq:Factorization}) and consider the effects
of decoherence on this initial state. For solid state implementations
we have that the most relevant source of noise is dephasing. In which
case we may model dephasing as a Krauss process with Krauss operators:
\begin{align}
\mathcal{K}_{i,\mathbb{I}} & =\sqrt{s_{i}}\mathbb{I}\nonumber \\
\mathcal{K}_{i,0} & =\sqrt{p_{i}}\left|0_{i}\right\rangle \left\langle 0_{i}\right|\nonumber \\
\mathcal{K}_{i,1} & =\sqrt{p_{i}}\left|1_{i}\right\rangle \left\langle 1_{i}\right|\label{eq:Krauss}
\end{align}
with $s_{i}=1-p_{i}=\exp\left(-\frac{t}{T_{2}^{*}}\right)$. Now the
total Krauss operator for the density matrix with $N$ spins is given
by: 
\begin{equation}
\mathcal{K}_{\left\{ \alpha\right\} }^{N}=\prod_{i=1}^{N}\mathcal{K}_{i,\alpha_{i}}\label{eq:Product}
\end{equation}
Where $\alpha_{i}=\mathbb{I},0,1$. Now we wish to compute the expectation
value of an arbitrary operator $O_{N}$ under Krauss evolution, we
have that: 
\begin{align}
\left\langle O_{N}\left(t\right)\right\rangle  & =\sum_{\left\{ \alpha\right\} }Tr\left[O_{N}\prod_{i=1}^{N}\mathcal{K}_{i,\alpha_{i}}\rho_{N}\left(t=0\right)\prod_{i=1}^{N}\mathcal{K}_{i,\alpha_{i}}^{\dagger}\right]\nonumber \\
 & =\sum_{\left\{ \alpha\right\} }Tr\left[\prod_{i=1}^{N}\mathcal{K}_{i,\alpha_{i}}^{\dagger}O_{N}\prod_{i=1}^{N}\mathcal{K}_{i,\alpha_{i}}\rho_{N}\left(t=0\right)\right]\label{eq:Krauss-1}
\end{align}
Now we have that 
\begin{align}
\mathcal{K}_{i,\mathbb{I}}^{\dagger}\sigma_{x}^{i}\mathcal{K}_{i,\mathbb{I}} & =s_{i}\sigma_{x}^{i},\:\mathcal{K}_{i,0}^{\dagger}\sigma_{x}^{i}\mathcal{K}_{i,0}=\mathcal{K}_{i,1}^{\dagger}\sigma_{x}^{i}\mathcal{K}_{i,1}=0\nonumber \\
\mathcal{K}_{i,\mathbb{I}}^{\dagger}\sigma_{y}^{i}\mathcal{K}_{i,\mathbb{I}}^{\dagger} & =s_{i}\sigma_{y}^{i},\:\mathcal{K}_{i,0}^{\dagger}\sigma_{y}^{i}\mathcal{K}_{i,0}=\mathcal{K}_{i,1}^{\dagger}\sigma_{y}^{i}\mathcal{K}_{i,1}=0\label{eq:Zero}
\end{align}
Now for initial states close to z-axis we have that 
\begin{align}
\left\langle \sigma_{x}^{i}\left(t\right)\right\rangle  & =Tr\left[s_{i}\sigma_{x}^{i}\rho_{N}\left(t=0\right)\right]=0\nonumber \\
\left\langle \sigma_{y}^{i}\left(t\right)\right\rangle  & =Tr\left[s_{i}\sigma_{y}^{i}\rho_{N}\left(t=0\right)\right]=0\label{eq:Zero-1}
\end{align}
Furthermore we have that: 
\begin{align}
 & \left\langle \sigma_{\alpha}^{i}\left(t\right)\sigma_{\beta}^{j}\left(t\right)\right\rangle \nonumber \\
 & =s_{i}s_{j}Tr\left[\sigma_{\alpha}^{i}\sigma_{\beta}^{j}\rho_{N}\left(t=0\right)\right]\nonumber \\
 & =s_{i}s_{j}\left\langle \sigma_{\alpha}^{i}\left(t=0\right)\sigma_{\beta}^{j}\left(t=0\right)\right\rangle \label{eq:Two_spin}
\end{align}
With $\alpha,\beta=x/y$ and $i\neq j$. Now we have that the expectation
value of:
\begin{widetext}
\begin{align}
\left\langle \left[\sum_{i=1}^{N}\left[\sigma_{x}^{i}\left(t\right)\cos\left(\theta\right)+\sigma_{y}^{i}\left(t\right)\sin\left(\theta\right)\right]\right]^{2}\right\rangle  & =N+\sum_{i\neq j}\left\langle \left[\sigma_{x}^{i}\left(t\right)\cos\left(\theta\right)+\sigma_{y}^{i}\left(t\right)\sin\left(\theta\right)\right]\left[\sigma_{x}^{j}\left(t\right)\cos\left(\theta\right)+\sigma_{y}^{j}\left(t\right)\sin\left(\theta\right)\right]\right\rangle \nonumber \\
 & =N+\sum_{i\neq j}s_{i}s_{j}\left\langle \left[\sigma_{x}^{i}\cos\left(\theta\right)+\sigma_{y}^{i}\sin\left(\theta\right)\right]\left[\sigma_{x}^{j}\cos\left(\theta\right)+\sigma_{y}^{j}\sin\left(\theta\right)\right]\right\rangle \label{eq:Time}
\end{align}
\end{widetext}

Now we know that at $t=0$ we must have that:
\begin{align}
 & \frac{N\xi^{2}}{\frac{1}{N}\sum_{i}\left\langle \sigma_{z}^{i}\right\rangle }=N\nonumber \\
 & +\sum_{i\neq j}\left\langle \left[\sigma_{x}^{i}\cos\left(\theta\right)+\sigma_{y}^{i}\sin\left(\theta\right)\right]\left[\sigma_{x}^{j}\cos\left(\theta\right)+\sigma_{y}^{j}\sin\left(\theta\right)\right]\right\rangle \label{eq:Squeezing-1}
\end{align}
\begin{align}
 & \Rightarrow\sum_{i\neq j}\left\langle \left[\sigma_{x}^{i}\cos\left(\theta\right)+\sigma_{y}^{i}\sin\left(\theta\right)\right]\left[\sigma_{x}^{j}\cos\left(\theta\right)+\sigma_{y}^{j}\sin\left(\theta\right)\right]\right\rangle \nonumber \\
 & =-N\left(P-\xi^{2}\right)\label{eq:Squeezing}
\end{align}
Where $P=\frac{1}{N}\sum_{i}\left\langle \sigma_{z}^{i}\right\rangle $.
This is true for a generic squeezed state. Furthermore we will assume
that $s_{i}=s_{j}=s=\exp\left(-\frac{t}{T_{2}^{*}}\right)$ then we
have that: 
\begin{align}
\frac{N\xi^{2}\left(t\right)}{P} & =\frac{N-N\left(P-\xi^{2}\right)s^{2}}{P}\nonumber \\
\xi^{2}\left(t\right) & =1-\left(P-\xi^{2}\right)s^{2}\nonumber \\
\frac{\xi^{2}\left(t\right)}{\xi^{2}} & =s^{2}+\frac{P-s^{2}}{\xi^{2}}\label{eq:Coherence}
\end{align}

\section{Variable coupling squeezing}\label{sec:Varaible-couplign-squeezing}

\subsection{Squeezing}\label{subsec:Squeezing}

We write:
\begin{equation}
U\left(t\right)=\prod_{i\neq j}\exp\left(-i\theta_{ij}\sigma_{x}^{i}\sigma_{x}^{j}\right)\label{eq:Unitary-1}
\end{equation}
with $\theta_{ij}=\theta_{ji}$. Now we have that: 
\begin{widetext}
\begin{align}
\sigma_{x}^{k}\left(t\right) & =\sigma_{x}^{k}\nonumber \\
\sigma_{y}^{k}\left(t\right) & =\prod_{j\neq k}\exp\left(2i\theta_{kj}\sigma_{x}^{k}\sigma_{x}^{j}\right)\sigma_{y}^{k}\prod_{i\neq k}\exp\left(-2i\theta_{ki}\sigma_{x}^{k}\sigma_{x}^{i}\right)=\sigma_{y}^{k}\prod_{j\neq k}\left(\cos\left[4\theta_{jk}\right]-i\sin\left[4\theta_{jk}\right]\left(\sigma_{x}^{j}\sigma_{x}^{k}\right)\right)\label{eq:Expansion}\\
\sigma_{z}^{k}\left(t\right) & =\prod_{j\neq k}\exp\left(2i\theta_{kj}\sigma_{x}^{k}\sigma_{x}^{j}\right)\sigma_{z}^{k}\prod_{i\neq k}\exp\left(-2i\theta_{ki}\sigma_{x}^{k}\sigma_{x}^{i}\right)=\sigma_{z}^{k}\prod_{j\neq k}\left(\cos\left[4\theta_{jk}\right]-i\sin\left[4\theta_{jk}\right]\left(\sigma_{x}^{j}\sigma_{x}^{k}\right)\right)
\end{align}
Now we write: 
\begin{equation}
\left\langle \sigma_{x}^{k}\left(t=0\right)\right\rangle =\left\langle \sigma_{y}^{k}\left(t=0\right)\right\rangle =0,\quad\left\langle \sigma_{z}^{k}\left(t=0\right)\right\rangle =P_{k}\label{eq:Polarization}
\end{equation}
Where $P_{i}$ is the polarization. Now we have that: 
\begin{equation}
\left\langle \sigma_{x}^{k}\left(t\right)\right\rangle =\left\langle \sigma_{y}^{k}\left(t\right)\right\rangle =0,\quad\left\langle \sigma_{z}^{k}\left(t=0\right)\right\rangle =P_{k}\prod_{j\neq k}\cos\left[4\theta_{jk}\right]\label{eq:Polarizations}
\end{equation}
Now we have that for $l\neq k$:
\begin{equation}
\left\langle \sigma_{x}^{k}\left(t\right)\sigma_{x}^{l}\left(t\right)\right\rangle =\left\langle \sigma_{x}^{k}\sigma_{x}^{l}\right\rangle =0\label{eq:zero}
\end{equation}
And:
\begin{align}
 & \left\langle \sigma_{y}^{k}\left(t\right)\sigma_{y}^{l}\left(t\right)\right\rangle \nonumber \\
 & =\left\langle \left[\sigma_{y}^{k}\prod_{j\neq k}\left(\cos\left[4\theta_{jk}\right]-i\sin\left[4\theta_{jk}\right]\left(\sigma_{x}^{i}\sigma_{x}^{k}\right)\right)\right]\left[\sigma_{y}^{l}\prod_{i\neq l}\left(\cos\left[4\theta_{il}\right]-i\sin\left[4\theta_{il}\right]\left(\sigma_{x}^{i}\sigma_{x}^{l}\right)\right)\right]\right\rangle \nonumber \\
 & =\left\langle \sigma_{y}^{k}\left(\cos\left[4\theta_{jk}\right]-i\sin\left[4\theta_{jk}\right]\left(\sigma_{x}^{l}\sigma_{x}^{k}\right)\right)\sigma_{y}^{l}\left(\cos\left[4\theta_{kl}\right]-i\sin\left[4\theta_{il}\right]\left(\sigma_{x}^{k}\sigma_{x}^{l}\right)\right)\right.\nonumber \\
 & \left.\prod_{j\neq k,l}\left[\left(\cos\left[4\theta_{jk}\right]-i\sin\left[4\theta_{jk}\right]\left(\sigma_{x}^{i}\sigma_{x}^{k}\right)\right)\left(\cos\left[4\theta_{il}\right]-i\sin\left[4\theta_{il}\right]\left(\sigma_{x}^{i}\sigma_{x}^{l}\right)\right)\right]\right\rangle \\
 & =\left\langle \sigma_{y}^{k}\sigma_{y}^{l}\prod_{j\neq k,l}\left(\cos\left[4\theta_{jk}\right]\cos\left[4\theta_{jl}\right]-\sin\left[4\theta_{jk}\right]\sin\left[4\theta_{jl}\right]\left(\sigma_{x}^{l}\sigma_{x}^{k}\right)\right)\right\rangle \nonumber \\
 & =-P_{l}P_{k}\prod_{j\neq k,l}\left(\cos\left[4\theta_{jk}\right]\cos\left[4\theta_{jl}\right]-\sin\left[4\theta_{jk}\right]\sin\left[4\theta_{jl}\right]\right)+P_{l}P_{k}\prod_{j\neq k,l}\left(\cos\left[4\theta_{jk}\right]\cos\left[4\theta_{jl}\right]+\sin\left[4\theta_{jk}\right]\sin\left[4\theta_{jl}\right]\right)\label{eq:yy_noise}
\end{align}

Furthermore: 
\begin{align}
 & \left\langle \sigma_{x}^{k}\left(t\right)\sigma_{y}^{l}\left(t\right)\right\rangle \nonumber \\
 & =\left\langle \sigma_{x}^{k}\sigma_{y}^{l}\left(\cos\left[4\theta_{lk}\right]-i\sin\left[4\theta_{lk}\right]\left(\sigma_{x}^{k}\sigma_{x}^{l}\right)\right)\prod_{i\neq l,k}\left(\cos\left[4\theta_{il}\right]-i\sin\left[4\theta_{il}\right]\left(\sigma_{x}^{i}\sigma_{x}^{l}\right)\right)\right\rangle \nonumber \\
 & =P_{l}\sin\left[4\theta_{kl}\right]\prod_{i\neq l,k}\cos\left[4\theta_{il}\right]\label{eq:Noise_II}
\end{align}
Therefore we have that:
\begin{equation}
\xi_{\theta}^{2}\equiv\frac{\left\langle \left[\sum_{i}\left(\cos\left(\theta\right)\sigma_{x}^{i}+\sin\left(\theta\right)\sigma_{y}^{i}\right)\right]^{2}\right\rangle }{\left\langle \sum_{i}\sigma_{z}^{i}\right\rangle }=\frac{\mathcal{A}}{\mathcal{B}}\label{eq:Ratio}
\end{equation}
with 
\begin{align}
\mathcal{A} & =N-\sum_{k\neq l}P_{l}P_{k}\sin^{2}\left(\theta\right)\prod_{j\neq k,l}\left(\cos\left[4\theta_{jk}\right]\cos\left[4\theta_{jl}\right]-\sin\left[4\theta_{jk}\right]\sin\left[4\theta_{jl}\right]\right)-\nonumber \\
 & +\sum_{k\neq l}P_{l}P_{k}\sin^{2}\left(\theta\right)\prod_{j\neq k,l}\left(\cos\left[4\theta_{jk}\right]\cos\left[4\theta_{jl}\right]+\sin\left[4\theta_{jk}\right]\sin\left[4\theta_{jl}\right]\right)+2\sin\left(2\theta\right)\sum_{k\neq l}P_{l}\sin\left[4\theta_{kl}\right]\prod_{i\neq l,k}\cos\left[4\theta_{il}\right]\nonumber \\
\mathcal{B} & =\sum_{k}P_{k}\prod_{j\neq k}\cos\left[4\theta_{jk}\right]\label{eq:a_b}
\end{align}
\end{widetext}

\subsection{Statistical properties}\label{subsec:Statistical-properties}

\subsubsection{Perfect polarization}\label{subsec:Perfect-polarization}

We write
\begin{equation}
p\left(\left\{ \theta_{ij}\right\} \right)=\mathcal{N}\prod_{i<j}\exp\left(-\alpha\left(\theta_{ij}-\theta_{0}\right)^{2}\right)\label{eq:Gaussian}
\end{equation}
with $\mathcal{N}=\left(\frac{\pi}{\alpha}\right)^{\frac{N\left(N-1\right)}{4}}$.
We also set $P_{i}=1$ for simplicity as polarizations are often good
and $P_{i}$ enter the expression for squeezing either linearly or
quadratically (not as a product on the order of $N$ times). We are
also assuming weak squeezing $\prod_{j\neq k}\cos\left[4\theta_{jk}\right]\cong1$.
As such we have that: 
\begin{widetext}
\begin{align}
\bar{\xi_{\theta}^{2}} & \equiv\int\prod_{i<j}d\theta_{ij}p\left(\left\{ \theta_{ij}\right\} \right)\xi_{\theta}^{2}=\nonumber \\
 & =1+\sin^{2}\left(\theta\right)\left(N-1\right)\left(\frac{\pi}{\alpha}\right)^{\frac{N-2}{2}}\int\prod_{j\neq k,l}d\theta_{jk}d\theta_{jl}\exp\left[-\alpha\left(\sum_{j}\left[\left(\theta_{jk}-\theta_{0}\right)^{2}+\left(\theta_{jl}-\theta_{0}\right)^{2}\right]\right)\right]\times\nonumber \\
 & \times\left[-\prod_{j\neq k,l}\cos\left[4\theta_{jk}+4\theta_{jl}\right]+\prod_{j\neq k,l}\cos\left[4\theta_{jk}-4\theta_{jl}\right]\right]\nonumber \\
 & +2\sin\left(2\theta\right)\left(N-1\right)\left(\frac{\pi}{\alpha}\right)^{\frac{N-1}{2}}\int\prod_{j}d\theta_{jl}\exp\left[-\alpha\left(\sum_{j}\left(\theta_{jl}-\theta_{0}\right)^{2}\right)\right]\sin\left[4\theta_{kl}\right]\prod_{j\neq l,k}\cos\left[4\theta_{jl}\right]\label{eq:First_step}
\end{align}
Then we have that: 
\begin{align}
\bar{\xi_{\theta}^{2}} & =1+\frac{\sin^{2}\left(\theta\right)}{2^{N-2}}\left(N-1\right)\left(\frac{\pi}{\alpha}\right)^{\frac{N-2}{2}}\int\prod_{j\neq k,l}d\theta_{jk}d\theta_{jl}\exp\left[-\alpha\left(\sum_{j}\left[\left(\theta_{jk}-\theta_{0}\right)^{2}+\left(\theta_{jl}-\theta_{0}\right)^{2}\right]\right)\right]\times\nonumber \\
 & \times\left[-\sum_{\gamma_{j}=\pm}\prod_{j}\gamma_{j}\left[\exp\left[i\gamma_{j}\left[4\theta_{jk}+4\theta_{jl}\right]\right]\right]+\sum_{\eta_{j}=\pm}\prod_{j}\eta_{j}\left[\exp\left[i\eta_{j}\left[4\theta_{jk}-4\theta_{jl}\right]\right]\right]\right]\nonumber \\
 & -i\frac{\sin\left(2\theta\right)}{2^{N-2}}\left(N-1\right)\left(\frac{\pi}{\alpha}\right)^{\frac{N-1}{2}}\int\prod_{j}d\theta_{jl}\exp\left[-\alpha\left(\sum_{j}\left(\theta_{jl}-\theta_{0}\right)^{2}\right)\right]\times\left[\sum_{\kappa_{k}=\pm1}\kappa_{k}\exp\left[i\kappa_{k}\left[4\theta_{kl}\right]\right]\right]\times\left[\sum_{\kappa_{j}=\pm1}\prod_{j}\exp\left[i\kappa_{j}\left[4\theta_{jl}\right]\right]\right]\label{eq:second_step}
\end{align}
Then we have that: 
\begin{align}
\bar{\xi_{\theta}^{2}} & =1+\frac{\sin^{2}\left(\theta\right)}{2^{N-2}}\left(N-1\right)\left[-\sum_{\gamma_{j}=\pm}\exp\left[2\alpha\sum_{j}\left[\left(\theta_{0}-\frac{2i\gamma_{j}}{\alpha}\right)^{2}-\theta_{0}^{2}\right]\right]\right.\nonumber \\
 & \left.+\sum_{\eta_{j}=\pm}\exp\left[\alpha\sum_{j}\left[\left(\theta_{0}-\frac{2i\eta_{j}}{\alpha}\right)^{2}+\left(\theta_{0}+\frac{2i\eta_{j}}{\alpha}\right)^{2}-2\theta_{0}^{2}\right]\right]\right]\nonumber \\
 & -i\frac{\sin\left(2\theta\right)}{2^{N-2}}\left(N-1\right)\left(\frac{\pi}{\alpha}\right)^{\frac{N-1}{2}}\sum_{\kappa_{k}=\pm1}\kappa_{k}\exp\left[2\alpha\left[\left(\theta_{0}-\frac{2i\kappa_{k}}{\alpha}\right)^{2}-\theta_{0}^{2}\right]\right]\sum_{\kappa_{j}=\pm1}\exp\left[2\alpha\sum_{j}\left[\left(\theta_{0}-\frac{2i\gamma_{j}}{\alpha}\right)^{2}-\theta_{0}^{2}\right]\right]\label{eq:Third_step}
\end{align}
Finally we obtain: 
\begin{align}
\bar{\xi_{\theta}^{2}} & =1+\frac{\sin^{2}\left(\theta\right)}{2^{N-2}}\left(N-1\right)\left[-\sum_{\gamma_{j}=\pm}\exp\left[-\sum_{j}\left[\frac{8}{\alpha}-8i\theta_{0}\gamma_{j}\right]\right]+\sum_{\eta_{j}=\pm}\exp\left[-\frac{8}{\alpha}\right]\right]\nonumber \\
 & -i\frac{\sin\left(2\theta\right)}{2^{N-2}}\left(N-1\right)\sum_{\kappa_{k}=\pm1}\kappa_{k}\exp\left[-\left[\frac{4}{\alpha}-4i\theta_{0}\kappa_{k}\right]\right]\times\sum_{\kappa_{k}=\pm1}\kappa_{j}\exp\left[-\left[\frac{4}{\alpha}-4i\theta_{0}\kappa_{j}\right]\right]\nonumber \\
 & =1-\sin^{2}\left(\theta\right)\left(N-1\right)\exp\left(-\frac{8\left(N-2\right)}{\alpha}\right)\left[\cos^{N-2}\left(8\theta_{0}\right)-1\right]+2\sin\left(2\theta\right)\left(N-1\right)\exp\left(-\frac{4\left(N-1\right)}{\alpha}\right)\sin\left(4\theta_{0}\right)\cos^{N-2}\left(4\theta_{0}\right)\label{eq:Last_step}
\end{align}
\end{widetext}

\subsubsection{Why the exponential factor in Eq. (\ref{eq:Last_step}) is negligible}\label{subsec:Why-the-exponential}

Now we have that 
\begin{equation}
\left\langle \left(\theta_{ij}-\theta_{0}\right)^{2}\right\rangle =\frac{1}{\alpha}=\kappa^{2}\theta_{0}^{2}\label{eq:variance}
\end{equation}
where $\kappa\ll1$ represents the fractional deviation from a perfect
squeezing Hamiltonian. Now for the regular squeezing we have that
the optimum $\theta_{0}$ is given by: 
\begin{equation}
\theta_{0}\sim\frac{1}{N^{2/3}}\Rightarrow\alpha\sim\frac{1}{N^{4/3}}\Rightarrow\exp\left(-\frac{N}{\alpha}\right)=1+O\left(\frac{1}{N^{1/3}}\right)\label{eq:Small}
\end{equation}
As such for large $N$ we recover he usual squeezing answer: 
\begin{align}
\bar{\xi_{\theta}^{2}} & =1-\sin^{2}\left(\theta\right)\left(N-1\right)\left[\cos^{N-2}\left(8\theta_{0}\right)-1\right]\nonumber \\
 & +2\sin\left(2\theta\right)\left(N-1\right)\sin\left(4\theta_{0}\right)\cos^{N-2}\left(4\theta_{0}\right)\label{eq:Correct}
\end{align}
\foreignlanguage{american}{This suppression of the inhomogeneity makes
sense as squeezing in the large $N$ limit results from the interactions
of many spins and as such inhomogeneities are averaged over because
of the large number of spins leading to the squeezing of any pair
of spins.}

\section{Finite polarization effects}\label{sec:Finite-polarization-effects}

We write $P_{l}=P_{k}=P$ and $\theta_{ij}=\theta_{0}=Jt$. Therefore
we have that:
\begin{widetext}
\begin{align}
\xi_{\theta}^{2} & \equiv\frac{\left\langle \left[\sum_{i}\left(\cos\left(\theta\right)\sigma_{x}^{i}+\sin\left(\theta\right)\sigma_{y}^{i}\right)\right]^{2}\right\rangle }{\left\langle \sum_{i}\sigma_{z}^{i}\right\rangle }=\label{eq:Ratio-1}\\
 & =\frac{1-\left(N-1\right)P^{2}\sin^{2}\left(\theta\right)\left[\cos^{N-2}\left(8\theta_{0}\right)-1\right]+2P\left(N-1\right)\sin\left(2\theta\right)\sin\left(4\theta_{0}\right)\cos^{N-2}\left(4\theta_{0}\right)}{P\cos^{N-2}\left[4\theta_{0}\right]}\\
 & \cong P^{-1}-\left(N-1\right)P\sin^{2}\left(\theta\right)\left[\cos^{N-2}\left(8\theta_{0}\right)-1\right]+2\left(N-1\right)\sin\left(2\theta\right)\sin\left(4\theta_{0}\right)\cos^{N-2}\left(4\theta_{0}\right)
\end{align}
Now we write $\sin^{2}\left(\theta\right)=\frac{1}{2}\left(1-\cos\left(2\theta\right)\right)$.
As such we have that: 
\begin{align}
\xi_{\theta}^{2} & =P^{-1}-\frac{\left(N-1\right)}{2}P\left[\cos^{N-2}\left(8\theta_{0}\right)-1\right]+\nonumber \\
 & +\left(N-1\right)\left[\frac{1}{2}P\left[1-\cos^{N-2}\left(8\theta_{0}\right)\right]\cos\left(2\theta\right)+2\sin\left(2\theta\right)\sin\left(4\theta_{0}\right)\cos^{N-2}\left(4\theta_{0}\right)\right]\nonumber \\
\Rightarrow\xi_{min}^{2} & =P^{-1}-\frac{\left(N-1\right)}{2}P\left[\cos^{N-2}\left(8\theta_{0}\right)-1\right]+\nonumber \\
 & -\left(N-1\right)\sqrt{\frac{P^{2}}{4}\left(1-\cos^{N-2}\left(8\theta_{0}\right)\right)^{2}+4\sin^{2}\left(4\theta_{0}\right)\cos^{2\left(N-2\right)}\left(4\theta_{0}\right)}\nonumber \\
 & \cong P^{-1}+\frac{\left(N-1\right)}{2}P\left[1-\cos^{N-2}\left(8\theta_{0}\right)\right]-\nonumber \\
 & -\frac{\left(N-1\right)}{2}P\left[1-\cos^{N-2}\left(8\theta_{0}\right)\right]\left(1+\frac{1}{2}\left(\frac{4\sin^{2}\left(4\theta_{0}\right)\cos^{2\left(N-2\right)}\left(4\theta_{0}\right)}{\frac{P^{2}}{4}\left(1-\cos^{N-2}\left(8\theta_{0}\right)\right)^{2}}\right)-\frac{1}{8}\left(\frac{4\sin^{2}\left(4\theta_{0}\right)\cos^{2\left(N-2\right)}\left(4\theta_{0}\right)}{\frac{P^{2}}{4}\left(1-\cos^{N-2}\left(8\theta_{0}\right)\right)^{2}}\right)^{2}+...\right)\nonumber \\
 & =P^{-1}\left[1-\frac{N-1}{2}\left(4\left(\frac{\sin^{2}\left(4\theta_{0}\right)\cos^{2\left(N-2\right)}\left(4\theta_{0}\right)}{\left(1-\cos^{N-2}\left(8\theta_{0}\right)\right)}\right)-32P^{-2}\left(\frac{\sin^{4}\left(4\theta_{0}\right)\cos^{4\left(N-2\right)}\left(4\theta_{0}\right)}{\left(1-\cos^{N-2}\left(8\theta_{0}\right)\right)^{3}}\right)+...\right)\right]\nonumber \\
 & =P^{-1}\left[1-2\left(N-1\right)\left(\frac{\sin^{2}\left(4\theta_{0}\right)\cos^{2\left(N-2\right)}\left(4\theta_{0}\right)}{\left(1-\cos^{N-2}\left(8\theta_{0}\right)\right)}\right)\left[1-8P^{-2}\left(\frac{\sin^{2}\left(4\theta_{0}\right)\cos^{2\left(N-2\right)}\left(4\theta_{0}\right)}{\left(1-\cos^{N-2}\left(8\theta_{0}\right)\right)^{2}}\right)\right]\right]\label{eq:Phase}
\end{align}
Now we have that: 
\begin{align}
\cos^{2\left(N-2\right)}\left(4\theta_{0}\right) & =\exp\left(2\left(N-2\right)\ln\left(\cos\left(4\theta_{0}\right)\right)\right)\nonumber \\
 & \cong\exp\left(2\left(N-2\right)\ln\left[1-8\theta_{0}^{2}\right]\right)\nonumber \\
 & \cong\exp\left(2\left(N-2\right)\left[\left(-8\theta_{0}^{2}\right)\right]\right)\nonumber \\
 & \cong1+2\left(N-2\right)\left[\left(-8\theta_{0}^{2}\right)\right]+32\left(N-2\right)^{2}\theta_{0}^{4}\nonumber \\
\cos^{N-2}\left(8\theta_{0}\right) & =\exp\left(\left(N-2\right)\ln\left(\cos\left(8\theta_{0}\right)\right)\right)\nonumber \\
 & \cong\exp\left(\left(N-2\right)\ln\left[1-32\theta_{0}^{2}\right]\right)\nonumber \\
 & \cong\exp\left(\left(N-2\right)\left[-32\theta_{0}^{2}\right]\right)\nonumber \\
 & \cong1-32\left(N-2\right)\theta_{0}^{2}+512\left(N-2\right)^{2}\theta_{0}^{4}+\frac{2^{12}}{3}\left(N-2\right)^{3}\theta_{0}^{6}\label{eq:Expansions}
\end{align}
 
\begin{align}
\frac{\sin^{2}\left(4\theta_{0}\right)\cos^{2\left(N-2\right)}\left(4\theta_{0}\right)}{\left(1-\cos^{N-2}\left(8\theta_{0}\right)\right)} & \cong\frac{1}{2\left(N-2\right)}\frac{\left[1-\left(N-2\right)16\theta_{0}^{2}+32\left(N-2\right)^{2}\theta_{0}^{4}\right]}{\left[1-\left(N-2\right)16\theta_{0}^{2}+\frac{128}{3}\left(N-2\right)^{2}\theta_{0}^{4}\right]}\cong\frac{1}{2\left(N-1\right)}\left[\left(1-\left(N-1\right)^{2}\frac{32}{3}\theta_{0}^{4}\right)\right]\nonumber \\
\frac{\sin^{2}\left(4\theta_{0}\right)\cos^{2\left(N-2\right)}\left(4\theta_{0}\right)}{\left(1-\cos^{N-2}\left(8\theta_{0}\right)\right)^{2}} & =\frac{1}{128\left(N-2\right)^{2}\theta_{0}^{2}}\cong\frac{1}{128\left(N-1\right)^{2}\theta_{0}^{2}}\label{eq:Simplifications}
\end{align}
This means that: 
\begin{align}
\xi_{min}^{2} & \cong P^{-1}\left[1-\left(1-\left(N-1\right)^{2}\frac{32}{3}\theta_{0}^{4}\right)\left[1-\frac{P^{-2}}{16\left(N-1\right)^{2}\theta_{0}^{2}}\right]\right]\nonumber \\
 & \cong P^{-1}\left[\frac{P^{-2}}{16N^{2}\theta_{0}^{2}}+\frac{32}{3}N^{2}\theta_{0}^{4}\right]+....\nonumber \\
 & =P^{-1}\left[\frac{P^{-2}}{16N^{2}J^{2}t^{2}}+\frac{32}{3}N^{2}J^{4}t^{4}\right]\label{eq:Final_polarization}
\end{align}
Now we have that 
\begin{equation}
\xi_{\theta}^{2}\cong\xi_{min}^{2}+\left(P^{-1}-\frac{\left(N-1\right)}{2}P\left[\cos^{N-2}\left(8\theta_{0}\right)-1\right]-\xi_{min}^{2}\right)\left(1-\cos\left(2\left[\theta-\theta_{min}\right]\right)\right)\label{eq:Final_formula}
\end{equation}
Where 
\begin{equation}
\tan\left(2\theta_{min}\right)=\frac{4\sin\left(4\theta_{0}\right)\cos^{N-2}\left(4\theta_{0}\right)}{P\left[1-\cos^{N-2}\left(8\theta_{0}\right)\right]}\Rightarrow\theta_{min}\cong8\theta_{0}=8Jt\label{eq:Simpler}
\end{equation}
Combining we obtain that: 
\begin{equation}
\xi_{\theta}^{2}\cong\xi_{min}^{2}+\left(P^{-1}+16\left(N-1\right)\left(N-2\right)P\theta_{0}^{2}-\xi_{min}^{2}\right)\left(1-\cos\left(2\left[\theta-8\theta_{0}\right]\right)\right)\label{eq:Final_answer}
\end{equation}
\end{widetext}

We furthermore note that 
\begin{align}
0 & =\frac{\partial}{\partial t}\left[P^{-1}\left[\frac{P^{-2}}{16N^{2}J^{2}t^{2}}+\frac{32}{3}N^{2}J^{4}t^{4}\right]\right]\nonumber \\
\Rightarrow0 & =-\frac{P^{-2}}{8N^{2}J^{2}t^{3}}+\frac{128}{3}N^{2}J^{4}t^{3}\nonumber \\
t & =\frac{3^{1/6}}{2^{5/3}}P^{-1/3}\frac{1}{JN^{1/3}}\nonumber \\
\xi_{min}^{2} & \sim P^{-7/3}N^{-4/3}\label{eq:Polarization-1}
\end{align}

\selectlanguage{american}%

\section{Argument why Eq. (\ref{eq:Commutator_zero}) is reasonable}\label{sec:Argument-why-Eq.}

We consider the case of a large number of spins. In which case we
have that for a term in the density matrix: 
\begin{equation}
\rho=\sum_{\alpha}P_{\alpha_{i}}\prod\sigma_{\alpha_{i}}^{i},\quad\alpha_{i}=\mathbb{I},x,y,z\label{eq:Density-1}
\end{equation}
Because the polarization is non-zero there is a large number $\sim N$
terms with $\alpha_{i}=x,y,z$. As such we have that 
\begin{equation}
\mathcal{L}_{2}\left(\rho\right)=-2\sum_{\alpha}P_{\alpha_{i}}\prod\sigma_{\alpha_{i}}^{i}\Gamma_{\parallel/\perp}\left(1-\delta_{\alpha_{i},\mathbb{I}}\right)\label{eq:Decoherence}
\end{equation}
contains a large number of terms. Now we have that any terms of the
form: 
\begin{equation}
\mathcal{L}_{1x}^{ij}\left(\rho\right):\sum_{i}\left(1-\delta_{\alpha_{i},\mathbb{I}}\right)\rightarrow\sum_{i}\left(1-\delta_{\alpha_{i},\mathbb{I}}\right)+\left(0/+1/-1\right)\label{eq:Change}
\end{equation}
where 
\begin{equation}
\mathcal{L}_{1}^{ij}\left(\rho\right)=J\sigma_{x}^{i}\sigma_{x}^{j}\label{eq:L_1}
\end{equation}
As such the rate of decoherence does not significantly change for
a typical term in Eq. (\ref{eq:Decoherence}). So we have that: 
\begin{equation}
\left[\mathcal{L}_{1x/y}^{ij},\mathcal{L}_{2}\right]\cong0\label{eq:Zero-1-1}
\end{equation}
which implies Eq. (\ref{eq:Commutator_zero}).


\begin{thebibliography}{10 (2008)}
\bibitem[1(1993)]{Kitagawa_1993} M. Kitagawa, and M. Ueda, Phys.
Rev. A 47, 5138 (1993).

\bibitem[2(1992)]{Wineland_1992} D. J. Wineland, J. J. Bollinger,
W. M. Itano, F. L. Moore, and D. J. Heinzen, Phys. Rev. A 46, R6797
(1992).

\bibitem[3(1994)]{Wineland_1994} D. J. Wineland, J. J. Bollinger,
W. M. Itano, and D. J. Heinzen, Phys. Rev. A 50, 67 (1994).

\bibitem[4(2001)]{Sorensen_2001} A. Sorensen, L. Duan, J. Cirac,
and P. Zoller, Nature 409, 63 (2001).

\bibitem[5(2001)]{Bigelow_2001}N. Bigelow, Nature 409, 27 (2001).

\bibitem[6(2009)]{Guehne_2009} O. Guehne and G. Toth, Phys. Rep.
474, 1 (2009).

\bibitem[7(2008)]{Polzik_2008} E. S. Polzik, Nature 453, 45 (2008).

\bibitem[8(2009)]{Cronin_2009} A. D. Cronin, J. Schmiedmayer, and
D. E. Pritchard, Rev. Mod. Phys. 81, 1051 (2009).

\bibitem[9(2011)]{Ma_2011} J. Ma, X. Wang, C. P.Sun and F. Nori,
Phys, Rep. 509, 89 (2011).

\bibitem[10(2008)]{Li_2008} Y. Li, Y. Castin, and A. Sinatra, Phys.
Rev. Lett. 100, 210401 (2008).

\bibitem[11(2009)]{Li_2009} Y. Li, P. Treutlein, J. Reichel, and
A. Sinatra, Eur. Phys. J. B 68, 365 (2009).

\end{thebibliography}
\end{document}